\documentclass[prl,twocolumn]{revtex4-1}
\usepackage{bm}
\usepackage{graphicx}
\usepackage{amsmath}
\usepackage{amssymb}
\usepackage[utf8]{inputenc}
\usepackage[T1]{fontenc}
\usepackage{color}
\usepackage{float}
\usepackage{cancel}
\usepackage{upgreek}
\usepackage{subfigure}
\usepackage[normalem]{ulem}
\usepackage[unicode=true,colorlinks=true,citecolor=blue]{hyperref}
\usepackage{placeins}
\usepackage{lipsum}
\usepackage{epsfig}
\usepackage{booktabs}
\usepackage[utf8]{inputenc}
\usepackage{wrapfig}
\newcommand{\nix}[1]{}

\renewcommand{\phi}{\varphi}
\renewcommand{\kappa}{\varkappa}

\renewcommand{\i}{\mathrm i}
\newcommand{\eps}{\varepsilon}

\newcommand{\pderiv}[2]{\frac{\partial #1}{\partial #2}}
\newcommand{\aver}[1]{\left \langle #1 \right \rangle}

\newcommand{\beq}{\begin{equation}}
\newcommand{\eeq}{\end{equation}}
\newcommand\beqa{\begin{eqnarray}}
\newcommand\eeqa{\end{eqnarray}}
\newcommand\bea{\begin{array}}
	\newcommand\eea{\end{array}}
\newcommand\ba{\begin{array}}
	\newcommand\ea{\end{array}}
\newcommand{\nn}{\nonumber}

\newcommand{\eV}{{\,\mathrm {eV}}}
\newcommand{\nm}{{\,\mathrm {nm}}}

\newcommand{\bmul}{\begin{multline}}
\newcommand{\emul}{{\end{multline}}}

\begin{document}

\title{Edge photocurrent in bilayer graphene due to inter-Landau-level transitions} 
%Cyclotron resonance induced edge photocurrent in bi-layer graphene

\author{S. Candussio$^1$, M.V. Durnev$^2$, S. Slizovskiy$^{3,4,6}$, T. Jötten$^1$, J. Keil$^1$, V.V. Bel'kov$^2$,   J. Yin$^3$,  Y. Yang$^3$, S.-K. Son$^7$,
%Seok-Kyun Son$^3$, 
A. Mishchenko$^3$, V. Fal'ko$^{3,4,5}$,  and S.D. Ganichev$^1$
}

\affiliation{$^1$Terahertz Center, University of Regensburg, 93040 Regensburg, Germany}
\affiliation{$^2$Ioffe Institute, 194021	St. Petersburg, Russia}
\affiliation{$^3$Department of Physics \& Astronomy, University of Manchester, Manchester M13 9PL, UK }
\affiliation{$^4$National Graphene Institute, University of Manchester, Manchester M13 9PL, UK }
\affiliation{$^5$Henry Royce Institute for Advanced Materials, Manchester, M13 9PL, UK}
\affiliation{$^6$St. Petersburg INP, Gatchina, 188300, St.Petersburg, Russia}
\affiliation{$^7$Department of Physics, Mokpo National University, Muan 58554, Republic of Korea}

\begin{abstract}
We report the observation of the resonant excitation of edge photocurrents in bilayer graphene subjected to terahertz radiation and a magnetic field. The resonantly excited edge photocurrent is observed for both inter-band (at low carrier densities) and intra-band (at high densities) transitions between Landau levels (LL). While the intra-band LL transitions can be traced to the classical cyclotron resonance (CR) and produce strong resonant features, the inter-band-LL resonances have quantum nature and lead to the weaker features in the measured photocurrent spectra.  The magnitude and polarization properties of the observed features agree with the semiclassical theory of the intra-band edge photogalvanic effect, including its Shubnikov-de-Haas oscillations at low temperatures. 
\end{abstract}

%\pacs{}
\maketitle

\section{Introduction}

The electronic and transport properties of graphene and  two-dimensional (2D) semiconductors are strongly dependent on properties of the systems edges. Prominent examples are  the quantum Hall regime~\cite{Klitzing1980,Halperin1982,DasSarma1997,Novoselov2006,CastroNeto2009},  the spin Hall effect in 2D topological insulators~\cite{Murakami2004,Kane2005,Bernevig2006,Koenig2007,Hasan2010,Qi2011},  or quantum valley Hall effects in chiral edge states~\cite{CastroNeto2009,Martin2008,Yao2009,Zhang2011,Jung2011,Vaezi2013}.  In general, the energy dispersion and masses of the edge charge carriers can substantially differ from those of the material bulk. Thus, experimental access to these fundamental material properties becomes important. Different experimental techniques have been employed to explore electronic properties of edges including measurements of the quantum Hall effect~\cite{Sanchez2017},  scanning tunneling microscopy (STM)~\cite{Nimi2006,Ritter2009,Yin2016}, spatially resolved Raman scattering~\cite{Casiraghi2009,Heydrich2010,Begliarbekov}, spin Hall effect in topological insulator systems~\cite{Koenig2007}, imaging of edge currents by  superconducting quantum interference devices (SQUID)~\cite{Nowack2013,Marguerite2019},  electron spin resonance~\cite{Slota2018}, high-resolution transmission electron microscopy (TEM)~\cite{Liu2009}, and  near-field infrared nanometre-scale microscopy (nanoscopy)~\cite{Ju2015,Fei2015,Jiang2016,Nikitin2016,Li2016}. In the last decade it has been demonstrated that an access to the spin/valley properties of the edge transport can be obtained employing edge photogalvanic currents excited by terahertz or infrared radiation~\cite{Karch2011,GlazovGanichev_review,Dantscher_2017,Luo2017,Ganichev2017,Xu2018,plank2019,Durnev2019,Candussio2020,Otteneder2020}. While detection and analysis of the cyclotron resonance at the edges vicinity would provide an informative and important tool to study such edge properties as the energy spectrum, cyclotron mass, carrier density (type), and relaxation times, this challenging goal is missing so far. These data can also be useful in the development of novel graphene-based THz detectors, see e.g. Refs~\cite{Vicarelli2012,Freitag2013,Koppens2014,Olbrich2016,Auton2016,Auton2017,Bandurin2018,Castilla2019,DelgadoNotario2020}.

Here, we report on the observation of 
resonantly enhanced edge photogalvanic currents which probe magnetotransport in nanometers-scale vicinity of the edges.
The effect is demonstrated in high-mobility $\upmu$m-scale bilayer graphene structures excited by THz radiation.
The peaks in the edge current are caused by absorption peaks at frequencies that match the energies of either intra-band or inter-band Landau level (LL) transitions subjected to selection rules and favorable LL occupancy. The stronger intra-band peaks correspond to classical cyclotron resonances (CRs), while the weaker inter-band resonances have no classical counterpart. Below, we loosely refer to all the observed peaks as ``CR'', making clear identification of their origin.

 The observed photocurrent flows along the sample edges and its direction is defined by the magnetic field polarity. 
Upon variation of the magnetic field the edge current behaviour follows the semiclassical shape of the CR in the absorbance. Our analysis of the positions and shape of the CR resonances shows that properties of the edges in studied bilayer graphene samples can be well described by parameters of the material bulk.  Depending on the carrier density the resonances are caused by either inter-band or intra-band transitions between lowest Landau levels at low carrier density or involves transitions between adjacent LL in the conduction band in the semiclassical regime at high carrier density. In the latter case the resonant edge photogalvanic current additionally exhibits Shubnikov-de-Haas effect related $1/B$-periodic magnetooscillations yielding information on the edge carrier density. Rotating the polarization plane of normally incident THz radiation we observed that the orientation of the radiations electric field vector only slightly affects the photocurrent amplitude. The developed microscopic theory of the edge photocurrent in semiclassical regime is based on Boltzmann’s kinetic equation  and is in a good agreement with the experiment. 
The effect described here is probed in bilayer graphene; we note, however, that it is of general nature and should be observable also in other two-dimensional  crystalline systems like monolayer graphene, 2D topological insulators, and transition metal dichalcogenide monolayers.

\section{Samples and methods}
\label{samples_methods}

\begin{figure}
	\centering
	\includegraphics[]{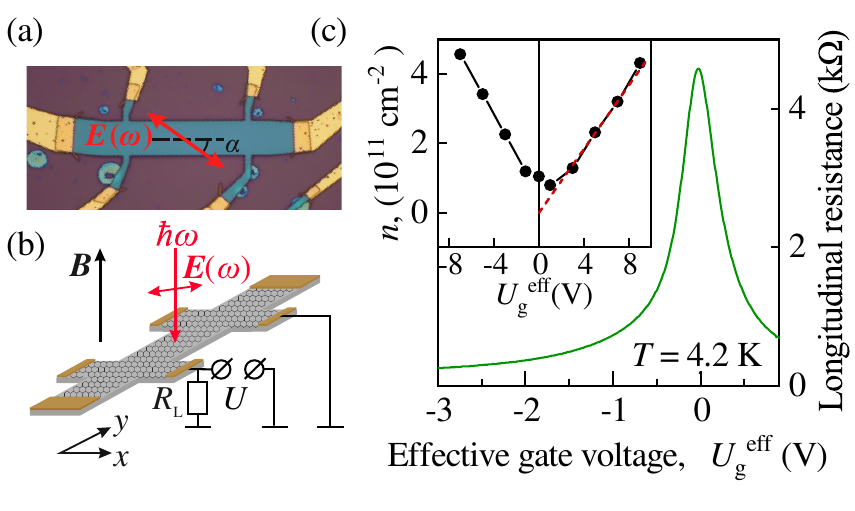}
	\caption{ Panel (a) shows a picture of the Hall bar structure. (b) Sketch of the measurement setup. The photovoltage is measured as a voltage drop $U$ over a load resistor $R_\text{L} = 10\, \text{M}\Omega$. (b) Longitudinal resistance versus the effective gate voltage. Corresponding carrier densities are shown in the inset. Dashed line is a linear fit after $n\,  [\text{cm}^{-2}] = 0.46 \times 10^{11} U_\mathrm{g}^{\mathrm{eff}}$~[V].}
	\label{fig_1}
\end{figure}

The experiments reported in this paper were carried out on Hall bar structures prepared from exfoliated bilayer graphene, encapsulated in hexagonal boron nitride and equipped with a back gate. We confirmed the bilayer nature of our graphene sample using Raman spectroscopy and low-temperature magnetotransport. Briefly, the spectra show a single sharp G peak at 1576~cm$^{-1}$ and a broad asymmetric 2D band with features typical for bilayer graphene \cite{LeeNovoselov2011}, for further details see Appendix~1.  $\text{SiO}_2$(290~nm)/Si serves as sample substrate and the contacts are made of Cr(3~nm)/Au. Figure \ref{fig_1}(a) shows a picture of the Hall bar structure, with dimensions $W/L = 9~\upmu\text{m}/31~\upmu \text{m}$. The gate voltage dependence of the longitudinal resistance presented in Figure~\ref{fig_1}(c) demonstrates a clear charge neutrality point (CNP). The curve was obtained in the absence of THz radiation applying a current of $I = 10^{-8}\text{A}$ modulated with a frequency of $12 \,\text{Hz}$. The CNP slightly shifts for different cool downs, therefore, we introduce an effective gate voltage $U_g^{\mathrm{eff}}=U_g - U_g^{\mathrm{CNP}}$. The carrier density $n$ as a function of the effective gate voltage obtained from magnetotransport experiments is presented in the inset of Fig.~\ref{fig_1}. The field effect mobility on the hole side is $\mu \approx 1.3\times 10^5$~cm$^2$/Vs and on the electron side it is $\mu \approx 1.5\times 10^5$~cm$^2$/Vs. More details on the transport and magnetotransport data are given in the Appendix~1.

\begin{figure}
	\centering
	\includegraphics[]{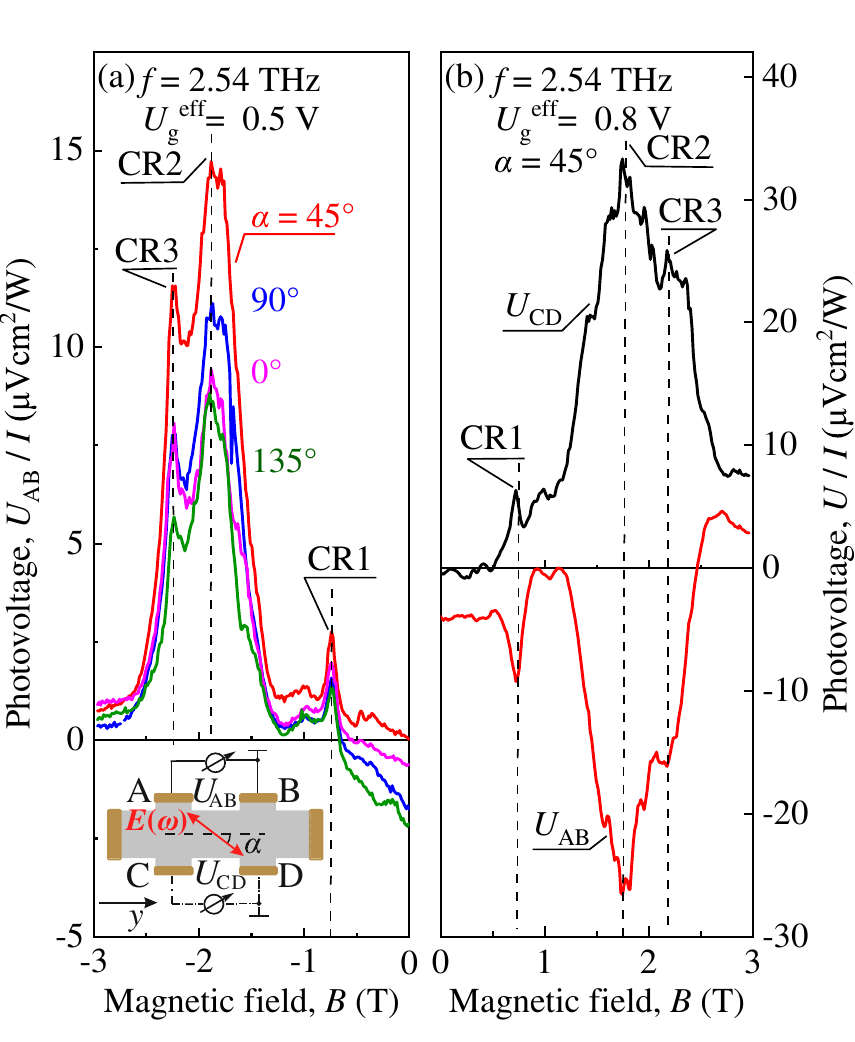}
	\caption{(a) Magnetic field dependence of the normalized photovoltage $U_{\rm AB}/ I \propto J_{\rm AB} /I$ 	measured for different azimuth angles $\alpha$. The data are presented for the radiation with $f=2.54$~THz, and, intensity $I\approx 2$~W/cm$^2$. The inset shows the experimental setup and defines the angle $\alpha$.  CR1, CR2, and CR3 label the resonance positions. (b)  Photovoltages $U_{\rm AB}$ and $U_{\rm CD}$  picked up from two contact pairs at the top (red line) and bottom (black line) sample edges aligned along the $y$-axis, see inset.}
	\label{fig_2}
\end{figure}

A continuous wave (cw) optically pumped molecular laser served as source of the normal incident THz radiation exciting the photocurrents with frequencies of  $f= 2.54$~THz (wavelength $\lambda = 118~\upmu$m, photon energy $\hbar\omega = 10.5$~meV) and $f=0.69$~THz ($\lambda = 432~\upmu$m, $\hbar\omega = 2.85$~meV)~\cite{Ganichev1982,Olbrich2013,Plank2016}. These laser lines were obtained using methanol and difluormethane as active media, respectively. A pyroelectric detector was used to control the radiation  power on the sample $P\approx 40$~mW.  The beam cross-section monitored by a pyroelectric camera had the Gaussian shape with spot sizes at the full width at half maximum (FWHM) of 1.5-2~mm, depending on the frequency. Consequently, the radiation intensity $I$ at the sample was about 2~W/cm$^2$. The laser used here emits linearly polarized radiation. In our setup the laser radiation electric field vector $\bm E$ was oriented along the $y$-axis, i.e. along the long side of the Hall bars. To rotate the electric field in respect to the edges we used crystal quartz $\lambda/2$-plates. The azimuth angle $\alpha$ is the angle between the $y$-axis and the radiation
electric field vector $\bm E$, see Fig.~\ref{fig_1}(a) and the inset of Fig.~\ref{fig_2}(a). 

The samples were placed in an optical cryostat and illuminated through $z$-cut crystal quartz window. All cryostat windows were covered by a thick black polyethylene film, which is transparent in the terahertz range, but prevents uncontrolled illumination of the sample with visible and infrared radiation. The magnetic field $B$ up to 5~T was applied normal to the graphene bilayer. Experiments were performed at temperature $T = 4.2$~K. A chopper placed in front of the laser modulated the radiation with a frequency of 40~Hz. The generated photocurrents $J \propto U$ were measured as a voltage drop $U$ over a load resistor of $R_L = 10\, \mathrm{M}\Omega$, see Fig.~\ref{fig_1}(a), applying a standard lock-in amplifier technique. 

\section{Resonant edge Photocurrent}

Upon excitation of the sample with linearly polarized radiation with $f=2.54$~THz and sweeping the magnetic field we observed three pronounced resonances at fields labeled in Fig.~\ref{fig_2} as CR1, CR2, and CR3. The data were detected between contacts located on the long Hall bar edges.  Figures~\ref{fig_2}(a) and (b) show two central observations: (i) the appearance of the resonances in the magnetic field dependence of the photoresponse and   (ii) the signal polarity is consistently opposite for opposite edges.  In addition, Fig.~\ref{fig_2}(a) reveals that the amplitude of the resonance depends on the orientation of the radiation electric field vector with respect to the edges. Positions of the resonances, their dependencies on the gate voltage, polarization, radiation frequency, and other parameters will be described and discussed later. First, we focus on the relative sign of the signals measured for the contact pairs AB and CD. The experimental setup is shown in the inset of Fig.~\ref{fig_2}(a). From the electric circuits used is seen that the positive signal picked up from the contacts CD and the negative one detected for the contacts AB indicate that the photocurrents along opposite edges flow in the opposite directions. This is only possible if the signal is generated at the edges and not in the graphene bulk (in the latter case current projections on the lines AB and CD would have the same direction). This observation, together with the fact that the signals for opposite edges have almost the same magnitudes, indicates that the signal stems from the photocurrent flowing along the  sample edges. The edge contribution $U_{\rm edge} \propto J_{\rm edge}$  can be obtained by subtracting the signals measured at the opposite sides, $U_{\rm edge} = (U_{\rm CD}-U_{\rm AB})/2 \propto (J_{\rm CD}-J_{\rm AB})/2$. 

%This figure and figure 2(c) reveal that the photoresponse is consistently of opposite signs for opposite edges and, therefore, originates from edge photocurrents.

\begin{figure}[h]
	\centering
	\includegraphics[width=\linewidth]{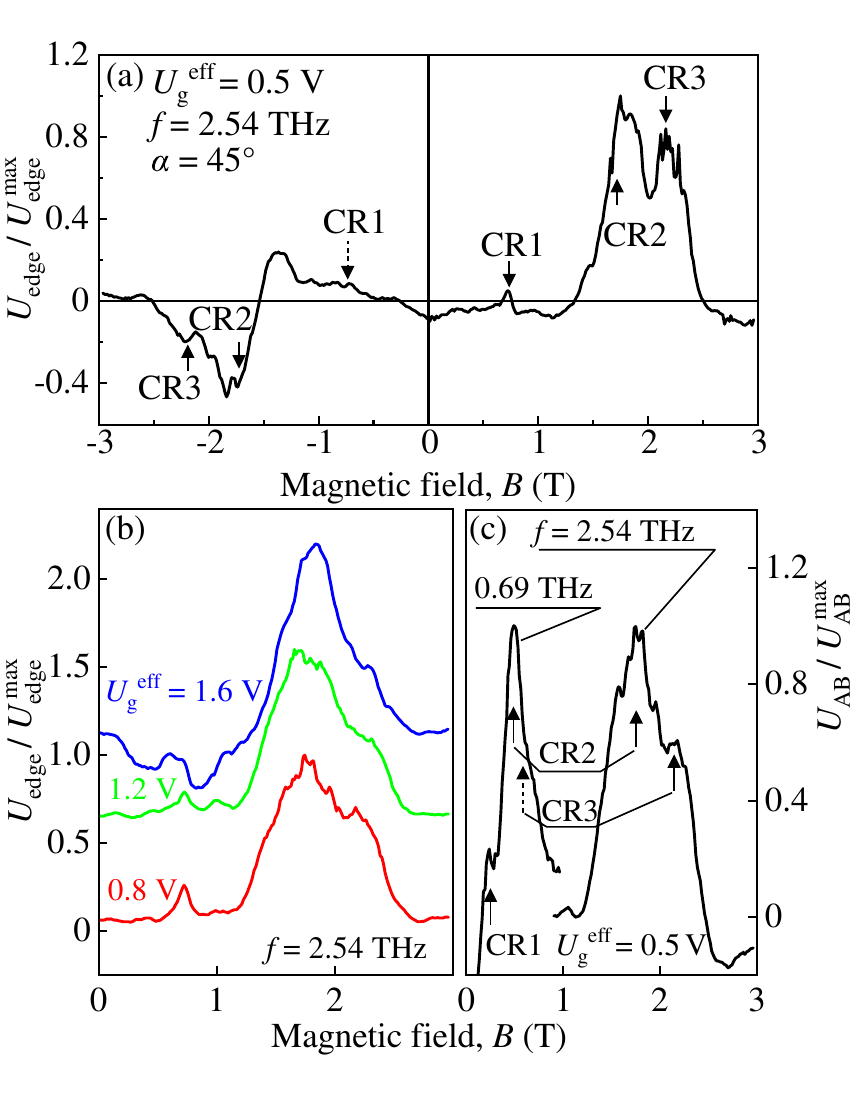}
	\caption{(a) Magnetic field dependence of the edge contribution of the photovoltage  $U_{\rm edge} = (U_{\rm CD}-U_{\rm AB})/2$ normalized to its maximal value $U^{\rm max}_{\rm edge}$  corresponding to the resonance CR2. The data are obtained for $f=2.54$~THz and $\alpha = 45^\circ$. CR1, CR2, and CR3 label the resonance positions. (b) Magnetic field dependence of the normalized edge photovoltage $U_{\rm edge}/U^{\rm max}_{\rm edge}$ measured for different effective gate voltages $U_g^{\mathrm{eff}}$. The curves are vertically shifted by 0.2 for visibility. (c) Magnetic field dependence of the  photovoltage $U_{\rm AB}$ normalized to its maximum value $U^{\rm max}_{\rm AB}$ corresponding to the resonance CR2. The data are presented for two radiation frequencies $f = 2.54$ and 0.69~THz. CR2 labels positions of the photovoltage maxima, whose position linearly shifts to a weak magnetic field with decreasing frequency according to $B_{\rm CR2}(\rm{2.54~THz}) / B_{\rm CR2}(\rm{0.69~THz}) = 2.54/0.69$.
}
	\label{fig_3}
\end{figure}

\begin{figure}
	\centering
	\includegraphics[]{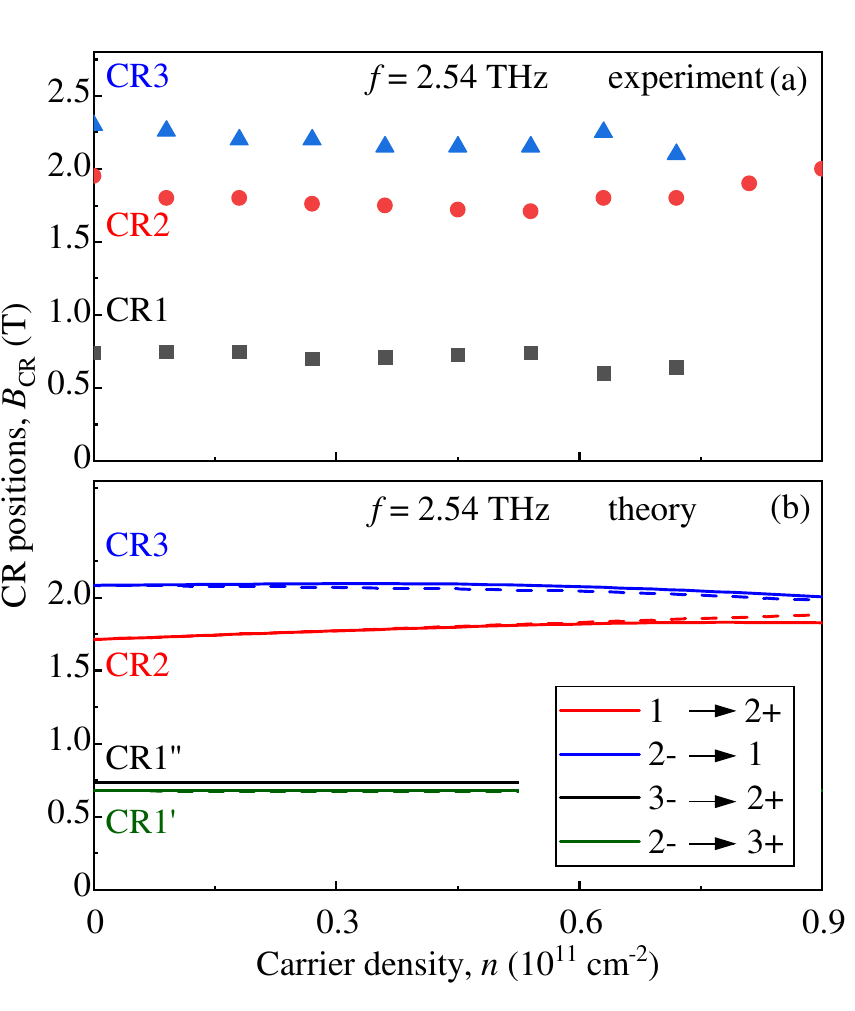}
	\caption{ Dependences of the cyclotron resonance positions $B_{\mathrm{CR1}}$,  $B_{\mathrm{CR2}}$, and $B_{\mathrm{CR3}}$ on the carrier density $n\,  [\text{cm}^{-2}] = 0.46 \times 10^{11} U_\mathrm{g}^{\mathrm{eff}}$~[V].  (a) data obtained from the photosignal resonances. (b) Values calculated from the allowed optical transitions between Landau levels are listed in the plot legend. Solid/dashed lines correspond to  valleys $ K^+ / K^-$. From the  position of CR2 corresponding to transitions within the conduction band we estimate the cyclotron mass $m_{\rm CR}$, which for zero density is $0.02\,m_0$. It increases with the increase of carrier density due to growth of the asymmetry parameter $\Delta$ with the gate voltage, see Eq.~(\ref{CRClassical}).}
	\label{fig_4}
\end{figure}

Figure~\ref{fig_3}(a) demonstrates that the resonance edge photosignals have opposite signs for opposite polarities of $B$, i.e.  is odd in magnetic field. Varying the Fermi level position by changing the back gate voltage we observed that at moderate gate voltages CR1 and CR3 become less pronounced in comparison to CR2, see Fig.~\ref{fig_3}(b), and  at even higher gate voltage these vanish.  Positions of CR1, CR2, and CR3 as a function of the carrier density are plotted in Fig.~\ref{fig_4}(a). When the carrier density changes from  $10^{10}$ to $10^{11}$~cm$^{-2}$, only weak shifts of the resonances are observed. 

Changing the radiation frequency from 2.54 to 0.69 THz results in the shift of CR2 and CR1 by about 3.7 times, which proves that the resonance positions scale linearly with $f$, see Fig.~\ref{fig_3}(c). Note that for $f=0.69$~THz the resonance CR3 vanish. As we will show below the resonance positions together with their behaviour upon variation of the carrier density and radiation frequency demonstrate that the resonances are caused by the cyclotron resonances involving transitions between valence and conduction bands (CR1 and CR3) and within the conduction band (CR2). 

\begin{figure}
	\centering
	\includegraphics[]{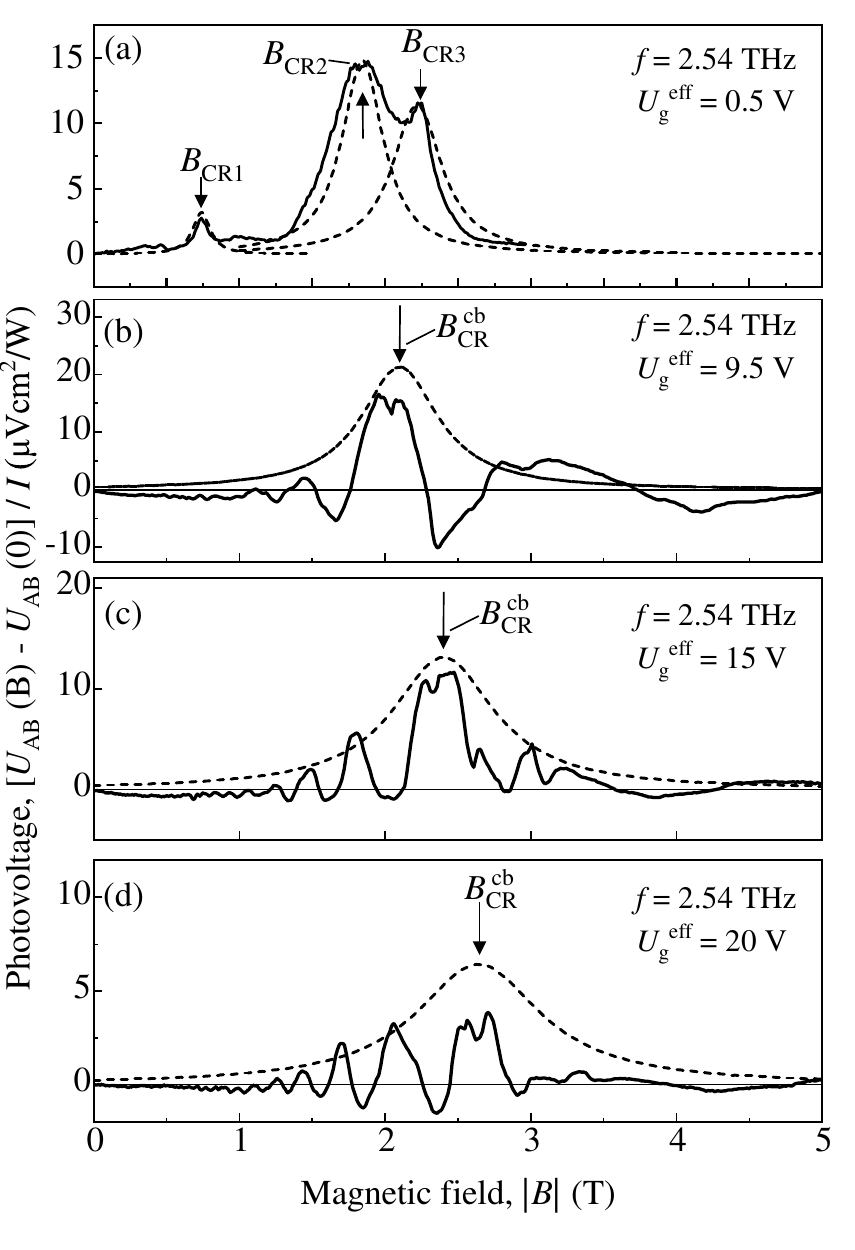}
	\caption{Magnetic field dependence of the photovoltage $[U_{\mathrm{AB}}(B) - U_{\mathrm{AB}}(0)]$. The data obtained for $\alpha = 45^{\circ}$ are given for different gate voltages increased from (a) $U_\mathrm{g}^\mathrm{eff}=0.5\,$V to (d) $U_\mathrm{g}^\mathrm{eff}=20\,$V. Dashed lines show fits after Eq.~(\ref{absorption}). The   values of $B_{\mathrm{CR1}}$ (for transitions $ 2- \to 3+$ and $3- \to  2+$ ) and $B_{\mathrm{CR3}}$ (for transitions $2- \to 1$) used for fits are calculated after Eqs.~(\ref{LLs}). The values of $B_{\rm CR2}$ and $B^{\rm cb}_{\rm CR}$ are calculated after Eq.~(\ref{CRClassical}) for transitions between neighboring Landau levels in the conduction band: $1 \to 2+,\; 2+ \to 3+,\; 4+ \to 5+,$ and $5+ \to 6+$ in panels (a), (b), (c), and (d), respectively.  Vertical arrows indicate the cyclotron resonance positions. The used transport broadening parameter was $\gamma = 1/\tau = 1.4$~ps$^{-1}$ and the radiative decay parameters $\Gamma$ are given in Tab.~\ref{tab1}.
	}
	\label{fig_5}
\end{figure}

The data discussed so far were obtained at low gate voltages (below 1.5~V) and, consequently, low carrier densities ($< 0.7\times 10^{11}\text{cm}^{-2}$). At higher carrier densities the analysis of the CR-induced photocurrent becomes more complicated, because the resonances are superimposed by the SdH-related oscillations in the photocurrent, see Fig.~\ref{fig_5}. The carrier densities calculated from the period of the $1/B$ oscillations of the photocurrent are the same as those calculated from the corresponding gate voltages. The figure shows, however, that the oscillating signal is strongly enhanced at magnetic fields of about 1.8 to 2.6~T. The observed enhancement  can be satisfactory described assuming a single resonance with a position depending on the carrier density. Our analysis below demonstrates that this behaviour is well described by the density dependence of the cyclotron resonance caused by optical transitions within the conduction band. Note that these results provide an additional evidence for the absence of resonances CR1 and CR3 at high carrier density. 

\begin{figure}[h]
 	\centering
 	\includegraphics[]{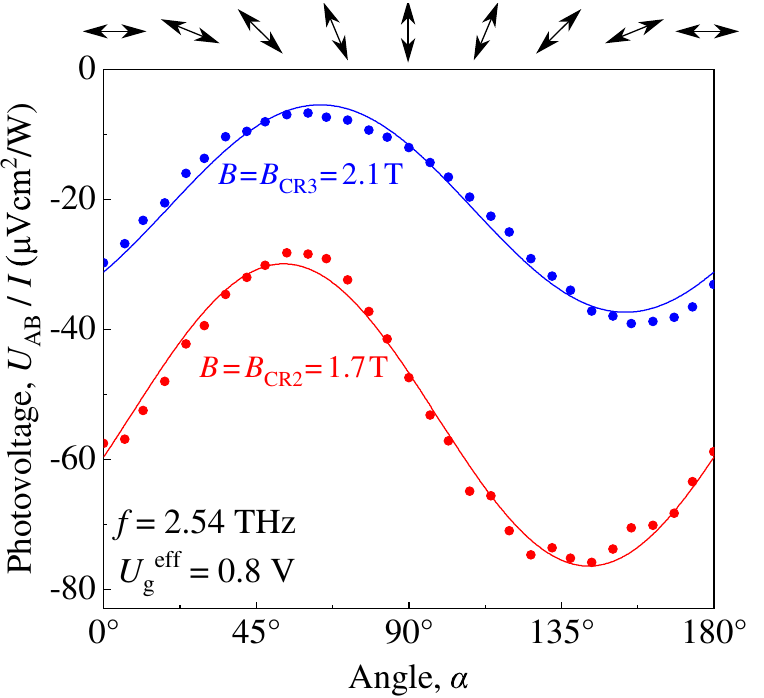}
 		 	\caption{ Dependence of the normailzed photovoltage $U_\mathrm{AB}/I$ on the azimuth angle $\alpha$. The data were obtained for	the magnetic fields $B_{\rm CR2}$ and $B_{\rm CR3}$. 	Solid lines are fits after Eq.~(\ref{fit-alpha}). Arrows on top illustrate states of polarization. Note that at  $\alpha = 0$ the electric field vector $\bm E$ is oriented parallel to the $y$-axis, i.e. to the long side of the Hall bar. }
 	\label{fig_6}
 \end{figure}

As addressed above, the magnitudes of the CR-induced photocurrents change upon rotation of the radiation electric field vector with respect to the edges, see Fig.~\ref{fig_2}(a). The dependence of the signal on the azimuth angle $\alpha$ can be well fitted by~\cite{Candussio2020,plank2019} 
\begin{equation} \label{fit-alpha}
U \propto J = J_{\rm L} \sin(2\alpha + \psi) + J_{\rm 0}, 
\end{equation}
see Fig.~\ref{fig_6} for contacts AB (upper edge). Here $J_{\rm L}$ and $J_{\rm 0}$ are the amplitudes of the polarization dependent and independent contributions, respectively. Note that for the CR-induced signals the phase shift $\psi$ is almost zero. Furthermore, while the signal strength varies with the azimuth angle, the sign of the resonance response remains the same and is defined by the magnetic field polarity only.

\section{Identification of cyclotron resonances}

To identify optical transitions responsible for the observed cyclotron resonances we obtain the energy dispersion of carriers in bilayer graphene as a function of the magnetic field and consider the corresponding selection rules. Bilayer graphene has a unit cell with 4 carbon atoms, $A_1, B_1, A_2, B_2$, where $A_1$ of the top layer lies above $B_2$ of the bottom layer. 
The dispersion is described with a Slonczewski-Weiss-McClure tight-binding 
model parameterized by \cite{SW58,McClure,Dresselhaus} parameters $\gamma_0 = 3.16 \eV$ (minus $A_i-B_i$ hoppings), $\gamma_1 = 0.35 \eV$ ($A_1 - B_2$ hopping), $\gamma_3=0.3 \eV$ (minus $B_1 - A_2$ hopping), $\gamma_4 = 0.14 \eV$ ($B_1 - B_2$ and $A_1 - A_2$ hoppings), $\Delta_{A_iB_i} = 0.05 \eV$ (sublattice energy difference $E_{A_1} - E_{B_1}$ and $E_{B_2} - E_{A_2}$) and top-bottom asymmetry parameter  $\Delta = \frac12 \left (E_{A_1} + E_{B_1} - E_{A_2} - E_{B_2}\right)$ that is induced by a vertical electric displacement field since the field of the bottom gate is not fully screened by the bottom graphene layer. Note that this parameter defines the gap. The resulting 4-band tight-binding model has been used to compute numerically the spectrum of Landau levels and the parameter $\Delta$ has been found from self-consistent Hartree approximation that is formulated as  \cite{McCannFalko06, BLGReview, SlizovskiyFalko}

\beq
\label{main2}
\Delta = \frac{4 \pi e^2 c_0}{\eps} \left[n_2  - \frac{\eps+1}{4} (n_1-n_2) \right],\nn
\eeq
where $c_0 = 0.335 \nm$ is the interlayer separation,  $\eps \approx 2.7$ accounts for electric polarizability of carbon orbitals in graphene and $n_{1,2}$  (constrained as $n = n_1 + n_2$) are the density of electrons at graphene layers that can be found from the wave functions of occupied Landau levels with phenomenological broadening.

To clarify the labeling of the Landau levels and give qualitative understanding of the results we note that the 4-band model can be reduced to an effective 2-band model with Hamiltonian \cite{BLGReview, McCannFalko06, 2BandH1} 
\begin{eqnarray} \nn
{\hat{H}}_{2} &=& {\hat{h}}_{0} + {\hat{h}}_{w} + {\hat{h}}_{4}
+ {\hat{h}}_{\Delta} + {\hat{h}}_{U} ,  \label{heff1} \\
{\hat{h}}_{0} &=&-\frac{1}{2m}\left(
\begin{array}{cc}
0 & \left( {\pi }^{\dag }\right) ^{2} \\  
{\pi ^{2}} & 0
\end{array} \right) , \nonumber \\
{\hat{h}}_{w} &=& v_{3}\left(
\begin{array}{cc}
0 & {\pi } \\
{\pi }^{\dag } & 0
\end{array}
\right) -  \frac{v_3 a}{4\sqrt{3}\hbar} \left( \begin{array}{cc}
0 & \left( {\pi }^{\dag }\right) ^{2} \\
{\pi ^{2}} & 0
\end{array} \right) , \nonumber
\\
{\hat{h}}_{4} &=& V_4 \left(
\begin{array}{cc}
\pi^{\dagger}\pi & 0 \\
0 & \pi \pi^{\dagger}
\end{array}
\right) ,  \ \ V_4 \equiv \frac{2v_0 v_4}{\gamma_1} + \frac{\Delta_{AB} v_0^2}{\gamma_1^2}  ,  \nonumber
\\
{\hat{h}}_{\Delta} &=& - \frac{\Delta}{2} \!\! \left[ \left(
\begin{array}{cc}
1 & 0 \\ 
0 & -1
\end{array}
\right)
- \frac{2v^2}{\gamma_1^2}
\left(
\begin{array}{cc}
\pi^{\dagger}\pi & 0 \\
0 & -\pi \pi^{\dagger} \\
\end{array}
\right) \right]  ,  \nonumber
\end{eqnarray}
where $\pi = \xi p_x + i p_y$, $v_i = \sqrt{3} a \gamma_i/2 \hbar$, $m = \gamma_1/(2 v_0^2)$,    $a$ is a lattice constant of graphene and $\xi = \pm 1$ denotes the $K^{\pm}$ valleys.  We note that ${\hat{h}}_{4}$ is responsible for electron-hole asymmetry of the spectrum (with holes being slightly more massive), ${\hat{h}}_{w}$ is a trigonal warping term and ${\hat{h}}_{\Delta}$ is a top-bottom asymmetry term that inevitably arizes in single-gated devices and is responsible for the gap between valence and conduction bands.

\begin{figure}[]
	\centering
	\includegraphics[width=\linewidth]{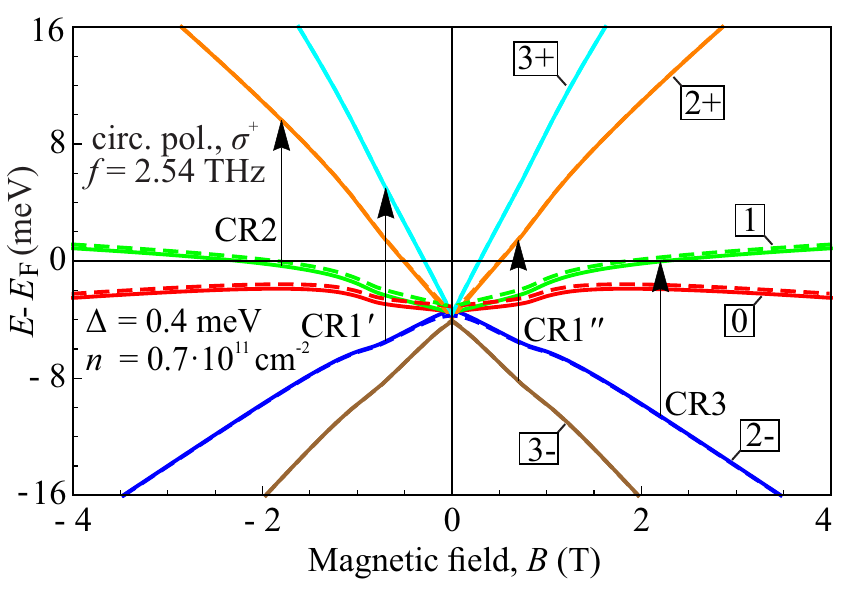}
	\caption{ Calculated Landau levels as a function of magnetic  field for gate induced doping $n= 0.7 \cdot 10^{11} {\rm cm}^{-2}$ corresponding to bilayer asymmetry gap $\Delta = 0.4\, $meV.  The levels are labeled according to Eqs.~(\ref{LLs}). Solid/dashed lines correspond to valleys $ K^+ / K^-$. Arrows indicate allowed optical transitions for right circularly polarized radiation ($\sigma^+$) with $f=2.54$ THz. Transitions labeled CR1$^\prime$, CR1$^{\prime \prime}$, CR2, and CR3 correspond to experimentally observed resonances, see, e.g., Fig.~\ref{fig_2}. Due to  close positions of  CR1$^\prime$ and CR1$^{\prime \prime}$ these resonances are not resolved in experiments, therefore,  in the corresponding experimental plots they are labeled as CR1. Note that for linear polarized radiation used in experiments the resonances plotted at positive (negative) magnetic fields appear also at negative (positive) $B$. } 
	\label{fig_7}
\end{figure}

In the standard basis of magnetic eigenstates  $\phi_l(x,y) = h_l(x/l_B - p_y l_B/\hbar) e^{i p_y y/\hbar}$, where $h_l$ are harmonic oscillator eigenstates and $l_B = \sqrt{\hbar c /eB}$,  the operators $\pi$ act as lowering operators $\pi \phi_l = - i \sqrt{2 l} (\hbar/l_B) \phi_{l-1}$ at the valley $K^+$ and as raising operators in the opposite valley. This allows us to find the spectrum of Landau levels analytically once we neglect the trigonal warping term. In the valley $K^+$ the spectrum is 
\beqa
E_{\pm l} &=& \frac{2 l-1}{l_B^2} \hbar^2  V_4 \pm \frac{\sqrt{l(l-1) + m^2 \left(V_4 + l_B^2 \Delta/(2 \hbar ^2)\right)^2}}{\hbar^{-2} l_B^2 m}, \nn \\
&&\psi_{l, \pm} \approx \frac{1}{\sqrt{2}} \left( \ba{c} \phi_l\nn \\ 
\pm \phi_{l-2} \ea \right),  \ \ l\ge 2 \nn \\
E_1 &=& \frac{\Delta}{2} + \frac{2 \hbar^2 V_4 }{l_B^2} , \ \  \psi_{1\, K_+} =  \left( \ba{c} \phi_1 \\ 0 \ea \right) \nn \\
E_0 &=& \frac{\Delta}{2} , \ \   \psi_{0,\, K_+} =  \left( \ba{c} \phi_0, \\ 0 \ea \right) \,. \label{LLs} 
\eeqa 
This clarifies the labeling of LLs as $3-, 2-, 0, 1, 2+, 3+…$, shown in Fig.~\ref{fig_7} ($+/-$ corresponds to Landau levels in conduction/valence band). 
To calculate the LL spectrum for the $K^-$ valley  $\Delta$ must be substituted by $-\Delta$.

The allowed optical transitions  are obtained from the selection rules following Ref.~[\onlinecite{AbergelFalko2007}]. For the circularly polarized radiation the allowed transitions are 
$$ (2-) \to (1) \; , \;  (3-) \to (2 \pm) \; , \;  (4-) \to (3 \pm)  ...  $$
when $\bm B$ is parallel to the direction of light angular momentum ($B>0$ in Fig.~\ref{fig_7})  or
$$... (3\mp) \to (4+) \; , \; (2 \mp) \to (3+) \; , \; (1) \to (2+)  $$
when $\bm B$ is anti-parallel to the direction of light angular momentum  ($B<0$ in Fig.~\ref{fig_7}).  The energy of $(1) \to (2+)$ and $(2-) \to (1)$ transitions are slightly valley-split when  $\Delta \neq 0$, while for other transitions the valley splitting is negligibly small. The transitions $(l-) \to ((l-1)-)$ at $l \gg 1$  correspond to classical CR transitions for holes,  transitions $(l+) \to ((l+1)+)$  are classical CR for electrons,  while the rest are the quantum inter-band transitions. Note that level $(0)$ does not participate in radiation emission/absorption in case of a small asymmetry parameter, see Ref.~\cite{AbergelFalko2007}. The selection rules need to be combined with the conditions that the initial (final) states are, at least partially, filled (empty). At low electron density and for a frequency of $f=2.54$ THz these yield  four resonances labeled CR1$^\prime$, CR1$^{\prime \prime}$, CR2, and CR3 in Fig.~\ref{fig_7}. The calculated positions correspond well to those detected in experiment, see, e.g., Fig.~\ref{fig_2}. Because CRs are excited by linearly polarized radiation the resonances in Fig.~\ref{fig_7} at positive (negative) magnetic field appear also for negative (positive) $B$. Note that in experiment the transitions CR1$^\prime$ and CR1$^{\prime \prime}$ are not resolved because they have very close positions. We also note that the difference in the CR2 and CR3 peak positions is due to  electron-hole asymmetry of the spectrum  (${\hat{h}}_{4}$ term in Hamiltonian), which results in a greater  mass of holes.  With the carrier  density increasing the final states for the inter-band transitions  in CR3,   CR1$^\prime$, and CR1$^{\prime \prime}$ become occupied and these resonances vanish. This is indeed observed in the experiment, see Fig.~\ref{fig_4}. The occupation of the final states is also the reason of disappearance of the CR3 resonance at substantial frequency reduction, see Fig.~\ref{fig_3}(c).

At high densities, the classical regime applies and the distance between the neighboring Landau levels $l$ and $l+1$ for $l\gg 1$, determining the position of CR, is given by 
\beqa \label{CRClassical} %\nn
E_{l+1} - E_{l} &\approx& \frac{2 \hbar^2 V_4}{l_B^2} \mathrm{sign}(l) \\ &+& \frac{\hbar^2}{l_B^2 m \sqrt{1 + m^2 \left(V_4 + l_B^2 \Delta/(2 \hbar ^2)\right)^2/l^2}}, \nn
\eeqa 
where we have to keep the term containing $\Delta$ since the latter grows almost linearly with increasing doping and hence is important even at large $l$. 
The values of resonance magnetic field $B_{\rm CR}$ are calculated using Eq.~\eqref{LLs} at low carrier density and Eq.~\eqref{CRClassical} at high carrier density corresponding to the semiclassical  regime. The calculated values of $B_{\rm CR}$
were then used to fit the data in Fig.~\ref{fig_5}.  The growth of $\Delta$ with gate voltage explains the reduction of LL separation, and, consequently, the growth of $B_{\rm CR}$ with increasing electron density.  
%
%The calculated positions of CR were used to fit the data in Fig. \ref{fig_5}. 
%The growth of $\Delta$ with gate voltage explains the reduction of the LL separation, and, consequently, the growth of CR magnetic field with increasing $n$.  

Below we show that at cyclotron resonance the edge	current flows within a strip with the width equal to the cyclotron diameter ($2 v /\omega_c$, where $v$ is the carrier velocity at the Fermi level). 
This value is around tens of nanometers and is already smaller than the free path length 	at experimental conditions.  Therefore, at CR we probe 	a few tens of nm wide region near the edge.  For low gate voltages, the doping near the edge is the same as in the bulk, so the CR position is determined by the bulk doping level. In general, at higher gate voltages (as used in Fig.~\ref{fig_5}), the doping may become notably inhomogeneous  near the edge of the graphene--back gate capacitor at the scale $d/\epsilon \approx 100\,$nm, where $d \approx 400\,$nm   is the distance to the bottom gate \cite{Silvestrov2008,Slizovskiy2018} and $\epsilon \approx 4$ is the dielectric constant of SiO$_2$ dielectric. This effect may lead to  broadening of the edge current region and to deviations of the CR position from that predicted from the bulk doping. However,  carrier densities obtained from the period of the photocurrent oscillations, Fig.~\ref{fig_5}, coincide with that calculated from the  corresponding gate voltages.

Our analysis below
%, see Secs.~\ref{sec:model} and \ref{sec:kinetic},  
demonstrates that the resonant behaviour of the observed edge photocurrents upon variation of the magnetic field follows the semiclassical shape of the CR in the absorbance, given by
%ID: Our analysis demonstrates that the decay of observed edge currents 	upon variation of the magnetic field on both sides of the CR follows	the expected semiclassical shape of the CR in the absorbance, given by
	%
	\begin{equation}
	\label{absorption}
	\eta  \propto \sum_{\pm}\frac{1}{(\Gamma + 1/\tau)^2 + (\omega \pm
		\omega_c)^2}\,.
	\end{equation}
Apart from the conventional transport width of the CR given by the 	momentum relaxation rate $1/\tau$, here we include an additional	electrodynamic contribution, $\Gamma = 2\pi e^2 n/c\, m_\text{CR}$, where $m_{\rm CR}$ is the cyclotron mass,	which accounts for a significant reflection of the radiation in the	vicinity of CR in high density samples~\cite{FalkoKhmelnitskii1989,Mikhailov2004,Herrmann2016}. The  radiative decay parameters $\Gamma$ calculated for different $U_\text{g}^\text{eff}$ are summarized in Tab.~\ref{tab1}. The curves calculated after the right-hand side of  Eq.~(\ref{absorption}) are shown by dashed lines in Fig.~\ref{fig_5} demonstrating a good agreement with the experimental 	data. In these plots we used the cyclotron electron mass $m_\text{CR}$ obtained from the calculated CR positions. 

The values of $B_{\rm CR}$ in the low-density regime  were calculated after Eq.~\eqref{LLs} for the transitions between the lowest Landau levels, see Fig.~\ref{fig_5}(a) and resonances $B_{\rm CR1}$, $B_{\rm CR2}$, and $B_{\rm CR3}$. In the high density regime $B_{\rm CR}$ was obtained from Eq.~\eqref{CRClassical} for the intra-band transitions between the neighboring Landau levels in the conduction band, see $B^{\rm cb}_{\rm CR}$ in Figs.~\ref{fig_5}(b), (c), and (d). 

At low electron density, see Fig.~\ref{fig_5}(a), the halfwidth of the CR detected in the edge vicinity is substantially larger than that expected for the radiative decay $\Gamma$, being for $U^\mathrm{eff}_g = 0.5$~V equal to 0.06 $\mathrm{ps}^{-1}$, see Tab.~\ref{tab1}. Therefore, the width of the resonance, in this case, is determined by the momentum relaxation time of the electrons in the edge channels $\tau$ \footnote{Note that the  values of the order of $\tau \approx 0.7$~ps corresponds well to those reported in Ref.~\cite{Cobaleda2014} obtained for the suspended bilayer graphene with the carrier mobility about $4\times10^4$~cm$^2$/Vs and the mass $0.03~\mathrm{m}_0$.}. The value of $\tau \approx 0.7$~ps, however, is half as much as expected for our samples with the carrier mobility of $\approx 1.5 \times 10^5$~cm$^2$/Vs and the cyclotron mass of $0.02~m_0$ (see Tab.~\ref{tab1}). We attribute the difference in the momentum relaxation times extracted from the transport measurements and from the CR data to the fact that the value of $\tau$ extracted from the edge photocurrent measurements may be shorter than that obtained in bulk magnetotransport experiments. At higher densities,  the radiative decay rate increases, see Tab.~\ref{tab1}, and becomes higher than the reciprocal momentum relaxation times. Thus, the width of the resonance detected at high carrier densities  is determined by the radiative decay and this value cannot be used to extract the momentum relaxation time at the edges.

While the overall agreement obtained without 	fitting parameters is encouraging, we should mention that the used 	semiclassical description becomes less accurate at low densities corresponding to transitions between low-lying Landau levels. Here the electrodynamic effects are negligible, but Landau quantization effect should be taken into account~\cite{Dmitriev2012}. In Fig.~\ref{fig_5}, one clearly distinguishes four CR lines	corresponding to different intra-band ($B_{\rm CR2}$ and $B^{\rm cb}_{\rm CR}$) and inter-band ($B_{\rm CR1}$  and $B_{\rm CR3}$) transitions between 	distinct Landau levels.
% Since all these resonances have similar shape	and width, the quasiclassical CR lineshape following Eq.~(\ref{absorption}) is shown only for the intra-band transition CR2.

\begin{table}[htbp]
	\centering
	\begin{tabular}{ccccccccc}
		\toprule
		$U_\text{g}^\text{eff}$ [V] &$n$ [$10^{11}\, \mathrm{cm}^{-2}$]& $B_\text{CR}$ [T]  & $m_\text{CR}/m_\text{0}$ & $\Gamma \, [\mathrm{ps}^{-1}]$ \\
		\midrule
		 0.5 & 0.23  &$B_{\rm CR2} = 1.8$          &  0.020 & 0.06 \\
		 9.5 & 4.37  &$B^{\rm cb}_{\rm CR} = 2.1$  &  0.023 & 1.00\\
		15   & 6.90  &$B^{\rm cb}_{\rm CR} = 2.4$  &  0.026 & 1.40 \\
		20   & 9.20  &$B^{\rm cb}_{\rm CR} = 2.7$  &  0.029 & 1.70 \\
		\bottomrule
	\end{tabular}
	\caption[Tabelle]{
Cyclotron resonance magnetic fields $B_{\rm CR2}$ and $B^{\rm cb}_{\rm CR}$, calculated cyclotron masses $m_\text{CR}/m_\text{0}$, and radiative decay parameter $\Gamma$ obtained for different gate voltages $U_\text{g}^\text{eff}$. The values together with the transport width of the CR given by the momentum relaxation rate $1/\tau = 1.4$~ps$^{-1}$ are used for calculating curves in  Fig.~\ref{fig_5}. } 
\label{tab1}
\end{table}

\section{Edge photocurrent at cyclotron resonance in classical regime} \label{sec:model}

In this section we discuss the model describing cyclotron resonances in the dc edge current. Edge photocurrent induced by terahertz radiation at zero and low magnetic fields far from the cyclotron resonance conditions was studied in bilayer graphene in our previous work~\cite{Candussio2020}. It was shown that the dc edge current is induced by the high-frequency electric field of the incident THz radiation due to the breaking of the parity symmetry at the edge. This phenomenon can be viewed as a dc conversion of the bulk ac electric current at the edge. Two possible microscopic mechanisms of such an ac-dc conversion were discussed. The first one is based on the alignment of the free carrier momenta by the high-frequency electric field and subsequent scattering of the carriers at the edge. The second one is related to the dynamic charge accumulation near the edge and synchronized charge motion along the edge. In line with Ref.~\cite{Candussio2020} the dc edge current observed in the present work at low magnetic field behaves as $J \propto \sin (2\alpha + \theta_B)$, where 
%$\alpha$ is the angle between the polarization vector of the high-frequency electric field and the edge, and 
$\theta_B$ is the phase shift induced by the applied magnetic field.

In the current work we present the theory of the edge photocurrent at high magnetic fields satisfying the cyclotron resonance conditions, when the driving frequency $\omega$ is close to the cyclotron frequency $\omega_c$. As seen in Figs.~\ref{fig_2} and \ref{fig_6}, the polarization and magnetic field dependencies of the edge photovoltage in the CR regime differ significantly from those at zero or low magnetic field addressed above. The sign of the voltage no longer depends on the orientation of the electric field with respect to the edge, see Ref.~\cite{Candussio2020}, but is defined by the direction of magnetic field.

\begin{figure}[h]
	\centering	\includegraphics[width=\linewidth]{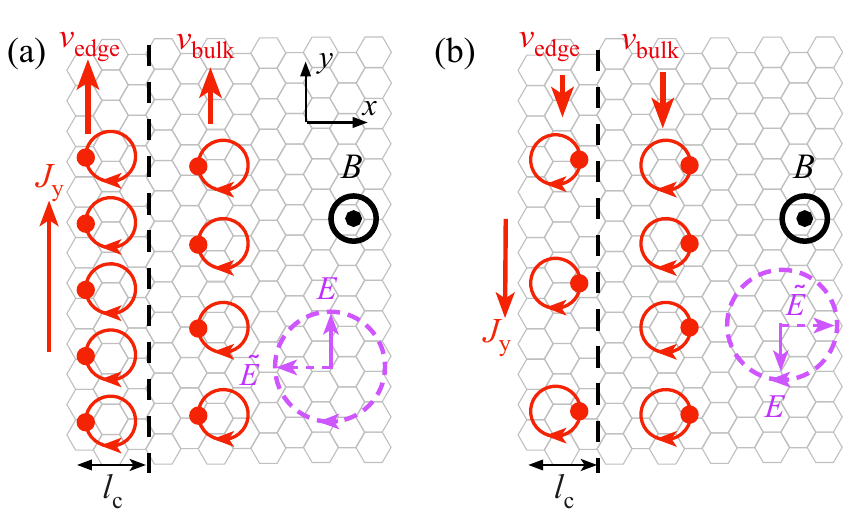}
	\caption{Schematic illustration of the edge current formation in the cyclotron resonance regime considering positively charged holes and circularly polarized radiation. Charge carriers move along cyclotron orbits. In the bulk the integral velocity of carriers $\bm v_{\rm bulk}$ rotates following rotating electric field $\bm E$ of radiation. The oscillating electric field leads to dynamic carrier accumulation and depletion at the edge, however the carrier density oscillates with a $\pi/2$ retardation as compared to $\bm E$ (one can say, that carrier density oscillations are driven by the electric field in the past $\tilde{\bm E}$). Due to the density oscillations the absolute value of the integral carrier velocity near the edge $\bm v_{\rm edge}$ is different at first [$E_y >0$, panel (a)] and second  [$E_y <0$, panel (b)] half-periods of the electric field oscillations, resulting in a net dc current flowing along the edge. 
		%The dc current is formed near the edge within the strip of the width equal to the diameter of the cyclotron orbit $l_c = 2v/\omega_c$.
%	The illustration is shown for positively charged holes and circularly polarized radiation.
	}
	\label{fig_mechanism}
\end{figure}

Such a behaviour can be explained by the following model, see Fig.~\ref{fig_mechanism}. Here, we discuss the model for positively charged holes, however its generalization to electrons is straightforward. At $\omega_c \tau \gg 1$ the holes in the 2D layer move along cyclotron orbits. When the frequency of the driving electric field coincides with the cyclotron frequency, the motion of holes becomes synchronized. The integral velocity of holes in the bulk of the 2D plane $\bm v_{\rm bulk}$ rotates following the rotating electric field $\bm E$ of the incident wave. The oscillating electric field also leads to periodic accumulation and depletion of holes at the edge, however, the hole density oscillates with a $\pi/2$ retardation as compared to electric field. This fact follows from the continuity equation $\partial n/\partial t + \partial i_x/ \partial x = 0$, where $n$ is the hole density and $i_x$ is the hole flux normal to the edge. One may say, that the oscillations of the hole density at the edge are driven by the electric field $\tilde{\bm E}$, which is shifted by a quarter of period in time with respect to $\bm E$. Due to holes accumulation and depletion, the integral hole velocity near the edge $\bm v_{\rm edge}$ differs from $\bm v_{\rm bulk}$. In the first half-period of the electric-field oscillations, when $E_y > 0$, the retarded field $\tilde{\bm E}$ points towards the edge leading to accumulation of holes at the edge and, consequently, to $|{\bm v}_{\rm edge}| > |{\bm v}_{\rm bulk}|$. At the second half-period, when $E_y < 0$, $\tilde{\bm E}$ points towards the bulk leading to depletion of holes at the edge, and during that time interval $|{\bm v}_{\rm edge}| < |{\bm v}_{\rm bulk}|$. Consequently in the vicinity of the edges such asymmetric oscillations of ${\bm v}_{\rm edge}$ result in a net dc current flowing along the edge.  This current is formed near the edge within a strip of the width equal to the diameter of the cyclotron orbit $l_c = 2v/|\omega_c|$. As seen in Fig.~\ref{fig_mechanism}, the direction of this current is determined by the direction of the cyclotron motion, and hence, by the sign of the magnetic field. The model suggests, that the increase of the edge current at the cyclotron resonance is related to the increase of $\bm v_{\rm bulk}$ and $\bm v_{\rm edge}$ due to resonant absorption.
% in the bulk.

The discussed mechanism describes the formation of the edge current at classical magnetic fields, when $\eps_F/\hbar \gg |\omega_c|$, where $\eps_F$ is the Fermi energy. It is valid for intra-band absorption at high carrier densities, see the observed $B^{\rm cb}_{\rm CR}$ resonance in Figs.~\ref{fig_5} (b-d). Resonances CR1 and CR3 are caused by inter-band transitions between valence and conduction band Landau levels and formation of the edge current in that case should be considered separately~\cite{plank2019}.

For microscopic description of the edge photocurrent we apply the Boltzmann approach, which is valid for the classical range of the driving field frequency $\omega \ll \eps_F/\hbar$, corresponding to intra-band absorption of the incident radiation. 
We consider a semi-infinite 2D gas of charge carriers occupying $x \geq 0$ half-space.
To derive the edge photocurrent we follow the procedure of Ref.~\cite{Candussio2020}. We further analyse the spectral and polarization dependencies of the calculated current in the vicinity of the cyclotron resonance.

The distribution function of the charge carriers (electrons or holes) $f(\bm p, x, t)$ is found from the Boltzmann kinetic equation
\begin{equation}
\label{kinetic_eq}
\pderiv{f}{t} + v_x \pderiv{f}{x} + e \left( \bm{\mathcal E}(x,t) + \frac{1}{c} \bm v \times \bm B \right) \cdot  \pderiv{f}{\bm p} = -\frac{f - \aver{f}}{\tau}\:,
\end{equation}
where $\bm{\mathcal E} (x,t) = \bm{\mathcal E}(x) \exp (-\i \omega t) + {\rm c.c.}$ is the total electric field in the sample consisting of the ac field of the incident wave $\bm E \exp (-\i \omega t) + {\rm c.c.}$ and the local screening field due to redistribution of charge, $\bm B \parallel z$ is the magnetic field, $\bm p$ and $\bm v = \bm p/m$ are the carrier momentum and velocity, respectively, $\aver{f}$ is the distribution function averaged over directions of $\bm p$, $\tau$ is the relaxation time, $e$ is the carrier electric charge and $m$ is the effective mass. In what follows we assume that $\tau$ is independent of carrier energy. The kinetic equation should be supplemented with a boundary condition at $x = 0$. Here we consider the specular reflection of carriers at the sample edge implying that the distribution function satisfies $f(p_x, p_y, 0) = f(-p_x, p_y, 0)$.

Equation~\eqref{kinetic_eq} is solved by expanding the distribution function in a series over the electric field $\bm E$ as $f(\bm p, x, t) = f_0(\bm p,x) + [f_1(\bm p,x) \exp(-\i \omega t) + {\rm c.c.}] + f_2(\bm p, x )$, where $f_0$ is the equilibrium distribution function, $f_1 \propto E$, and $f_2 \propto EE^*$~\cite{Karch2011}.
The density of the dc edge current is determined by the time-independent correction $f_2(\bm p, x)$ and reads
\begin{equation}
j_y(x) = 4e \sum \limits_{\bm p} v_y f_2(\bm p,x)\:,
\end{equation}
where the factor 4 accounts for the spin and valley degeneracy.
Using Eq.~\eqref{kinetic_eq} one can present $j_y(x)$ as~\cite{Candussio2020}
\begin{equation}
\label{jy}
j_y(x) = -4e\tau \sum \limits_{\bm p} v_x v_y \pderiv{f_2}{x} + 4\frac{e^2 \tau}{m} \sum \limits_{\bm p} (E_y^* f_1 + E_y f_1^*)\:.
\end{equation}

To calculate $j_y(x)$ using Eq.~\eqref{jy} one needs to find $f_1(\bm p,x)$ and $f_2(\bm p,x)$ from Eq.~\eqref{kinetic_eq}, which can be done numerically. However, as shown in Ref.~\cite{Candussio2020}, the total current flowing along the edge $J_y = \int_0^{\infty} j_y(x) dx$ is found analytically and can be expressed in terms of the conductivity tensor.
Substituting the expression for the conductivity tensor in
Eq.~(13) of Ref.~\cite{Candussio2020} and setting $\tau_1 = \tau_2 \equiv \tau$ we obtain
\begin{multline}
\label{Jy}
J_y =  \frac{e\tau^2 \sigma_0}{m} \left[ \mathcal A |\bm E|^2 + \i \mathcal B (E_x E_y^* - E_y E_x^*) \right. \\
\left. + \mathcal C (E_{x}E_{y}^* + E_{y}E_{x}^*) + \mathcal D (|E_{x}|^2 - |E_{y}|^2) \right]\:,
\end{multline}
where
\begin{eqnarray}
\label{ABCD}
\mathcal A &=& \frac{2 \omega_c \tau}{1 + 2(\omega^2 + \omega_c^2) \tau^2 + (\omega^2 - \omega_c^2)^2 \tau^4}\:, \\
\mathcal B &=& - \frac{1 + (\omega^2 + \omega_c^2) \tau^2}{\omega \tau \left[1 + 2(\omega^2 + \omega_c^2) \tau^2 + (\omega^2 - \omega_c^2)^2 \tau^4 \right]}\:, \nonumber \\
\mathcal C &=& -  \frac{1 + (\omega^2 - 5 \omega_c^2) \tau^2}{(1 + 4 \omega_c^2 \tau^2)[1 + 2(\omega^2 + \omega_c^2) \tau^2 + (\omega^2 - \omega_c^2)^2 \tau^4]}\:, \nonumber \\
\mathcal D &=& \frac{2\omega_c \tau \left[ 2 + (\omega^2 - \omega_c^2)\tau^2 \right]}{(1 + 4 \omega_c^2 \tau^2)[1 + 2(\omega^2 + \omega_c^2) \tau^2 + (\omega^2 - \omega_c^2)^2 \tau^4]}\:. \nonumber
\end{eqnarray}
Here $\sigma_0$ is the conductivity of 2D electron (or hole) gas at zero magnetic field, and $\omega_c = e B_z/(m_{\rm CR}c)$ is the cyclotron frequency. As follows from Eq.~\eqref{ABCD}, the dc edge current in presence of a magnetic field is induced by linearly polarized radiation with nonzero terms $E_{x}E_{y}^* + E_{y}E_{x}^*$ or $|E_{x}|^2 - |E_{y}|^2$, circularly polarized radiation with nonzero $\i (E_x E_y^* - E_y E_x^*)$, and also unpolarized radiation.

Let us now consider the behaviour of the edge photocurrent~\eqref{Jy} in the vicinity of the cyclotron resonance, i.e. in the frequency range $|\omega - |\omega_c|| \ll |\omega_c|$ and $|\omega_c| \tau \gg 1$.
For linearly polarized radiation the edge current is determined by the first, third, and fourth terms in Eq.~\eqref{Jy}. The cyclotron resonance occurs for both directions of magnetic field at $B_z = \pm B_{\rm CR}$, where $B_{\rm CR} = m_{\rm CR} \omega c/|e|$. 
Near the resonance the edge current given by Eqs.~\eqref{Jy} and \eqref{ABCD} is simplified to
\begin{multline}
\label{Jy_lin}
J_y^{\rm lin} =  \frac{{\rm sign}(B_z) c \tau  \eta I}{2 B_{\rm CR}}  \\
\times \left[ 1 + \frac{\sqrt{1 + (\omega - |\omega_c|)^2 \tau^2}}{2 |\omega_c| \tau} \sin (2 \alpha + \theta_B) \right]\:.
\end{multline}
Here $\alpha$ is the electric field angle with respect to the edge,
\begin{equation}
\label{absorp}
\eta = \frac{2\pi \sigma_0}{n_\omega c}  \frac{1}{1 + (\omega - |\omega_c|)^2 \tau^2}\:
\end{equation}
is the absorbance of the 2D layer in the vicinity of cyclotron resonance and far from the edge for non-polarized radiation, $I = c n_\omega |\bm E|^2/(2\pi)$ is the intensity of the radiation, and $n_\omega$ is the refractive index of the dielectric medium surrounding bilayer graphene. The phase shift $\theta_B$ is determined from
\begin{equation}
\label{theta}
\tan \theta_B = \frac{1 + |\omega_c| (\omega - |\omega_c|) \tau^2}{\omega_c \tau}\:.
\end{equation}

The sign, magnitude, and polarization dependence of the edge dc current given by Eqs.~\eqref{Jy_lin} and \eqref{absorp} are in agreement with experimental observations, shown in Figs.~\ref{fig_2}, \ref{fig_3}, and \ref{fig_6}. Equations~\eqref{Jy_lin} and \eqref{absorp}
show that  the edge dc current at cyclotron resonance is proportional to the absorbed energy. Therefore, an increase of the edge current magnitude at cyclotron resonance is due to the increase of absorption. The direction of the edge current is determined by the direction of magnetic field in agreement with experimental data in Figs.~\ref{fig_2} and \ref{fig_3}, which show the change of the photovoltage sign for opposite directions of the magnetic field. Interestingly, the direction of the dc current is the same for electrons and holes. These characteristic features are similar to those observed for the edge photocurrent at strong magnetic field under conditions of the quantum Hall effect~\cite{plank2019}.

As follows from Eq.~\eqref{Jy_lin}, the edge current induced by linearly polarized radiation consists of a polarization-independent contribution and a contribution $\propto \sin(2\alpha + \theta_B)$ that is sensitive to the orientation of the electric field. The second contribution $\propto |\omega_c| \tau$ is less than the first one. 
The phase shift $\theta_B$, given by Eq.~\eqref{theta}, vanishes at the resonance so that the polarization dependent part of $J_y^{\rm lin}$ behaves as $\sin 2\alpha$. Weak dependence of the edge current on polarization and $\sin 2\alpha$ behaviour are in line with experimental observations shown in Fig.~\ref{fig_6}.
Far from resonance, when $|\omega - |\omega_c|| \tau \gg 1$, we have $|\theta_B| = \pi/2$, so that the polarization dependent part of $J_y^{\rm lin}$ behaves as $\cos 2\alpha$.

\begin{figure}[h]
	\centering
	\includegraphics[width=\linewidth]{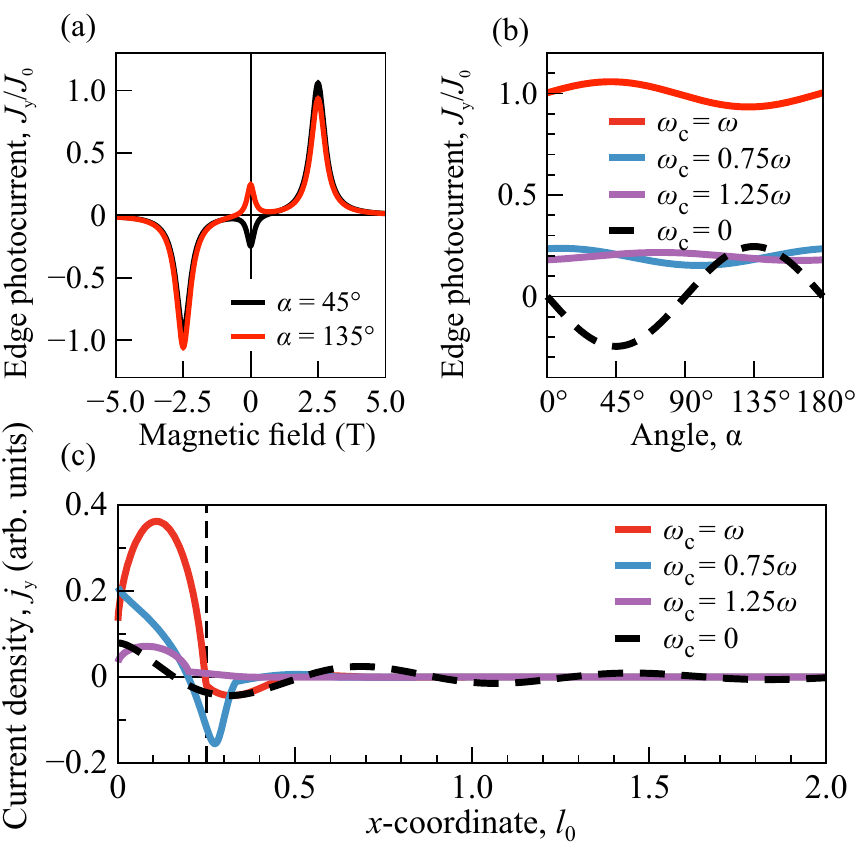}
	\caption{ (a) Edge photocurrent calculated after Eqs.~\eqref{Jy} and \eqref{ABCD} for linearly polarized radiation with $(E_x E_y^* + E_y E_x^*)/|\bm E|^2 = \pm 1$. Cyclotron resonances occur at $B_z = \pm 2.4$~T. (b) Dependence of the edge current on the azimuthal angle at different magnetic fields -- in resonance, $\omega_c = \omega$ (red line), and out of resonance, $\omega_c = 0.85~\omega$, $\omega_c = 1.15~\omega$. The curve at zero magnetic field is shown by a dashed line for comparison. (c) Distribution of the edge current density over $x$-coordinate for circularly polarized incident radiation. The dashed vertical line shows the diameter of the cyclotron orbit $l_c = 2 l_0/(\omega_c \tau)$ at resonance, $l_0$ is the mean free path length. The parameters used are: $\omega/(2\pi) = 2.54$~THz, $\tau = 0.7$~ps, $B_{\rm CR} = 2.4$~T, and $\omega \tau = 11$, $J_0 = c \tau \eta I/(2 B_{\rm CR})$.
	}
	\label{fig_Jy_theory}
\end{figure}

For circularly polarized radiation the edge photocurrent is determined by the first two terms in Eq.~\eqref{Jy}.
The cyclotron resonance occurs for a particular direction of the magnetic field at $B_z = - (e/|e|)P_{\rm circ} B_{\rm CR} $, where $P_{\rm circ} = \i (E_x E_y^* - E_y E_x^*)/|\bm E|^2$, $P_{\rm circ} = \pm 1$ for right (left)-hand circularly polarized radiation. Near the resonance Eq.~\eqref{Jy} yields
\begin{equation}
\label{Jy_circ}
J_y^{\rm circ} =  \frac{ {\rm sign}(B_z) c \tau  \eta I}{B_{\rm CR}} \:.
\end{equation}

Figure~\ref{fig_Jy_theory}(a) shows the edge photocurrent calculated after Eqs.~\eqref{Jy} and \eqref{ABCD} as a function of the magnetic field. The calculations were done for two linear polarizations $(E_x E_y^* + E_y E_x^*)/|\bm E|^2 = \pm 1$, corresponding to $\alpha = 45^\circ$ and $135^\circ$. Parameters used in calculations are the same as in calculations of the dashed curve in Fig.~\ref{fig_5}(c), see $U_{\rm g}^{\rm eff} = 15$~V in Tab.~\ref{tab1}. Sharp cyclotron resonances in the edge current occur at $B_z = \pm 2.4$~T with magnitude, which is almost independent of the polarization, and sign, which is determined by the direction of magnetic field. These results are in good agreement with experimental observations, see Figs.~\ref{fig_2}, \ref{fig_3}, and \ref{fig_5}.
By contrast, the directions of the dc current at $B_z = 0$ is opposite for $\alpha = 45^\circ$ and $135^\circ$, see Ref.~\cite{Candussio2020}. The dependence of $J_y$ on $\alpha$ is shown in Fig.~\ref{fig_Jy_theory}(b) for different values of $B_z$ corresponding to the resonances CR2 and CR3.
%exact resonance, out-of-resonance and zero magnetic field. It is seen, 
Note that the dependencies are presented for low densities corresponding to CRs in the quantum mechanical limit,  because at higher densities the  SdH-like oscillations make it difficult to extract the CR's amplitude and, consequently, do not allow us to analyze its polarization dependence.  Nevertheless, it is clear that the polarization dependence of the edge photocurrent at CR is indeed very weak. Its polarization dependent part follows $\sin 2\alpha$ dependence, whereas out of the resonance the polarization dependence of $J_y$ changes to $\cos 2\alpha$ (see Fig.~\ref{fig_Jy_theory}). This is in line with Eq.~\eqref{Jy_lin} and describes the experimental results, see Fig.~\ref{fig_6}. The distribution of the edge photocurrent density $j_y(x)$, calculated after Eq.~\eqref{jy} with $f_1(\bm p, x)$ found numerically from Eq.~\eqref{kinetic_eq}, is shown in Fig.~\ref{fig_Jy_theory}(c). The results are obtained for circularly polarized radiation when the first term is Eq.~\eqref{jy} vanishes. It is seen that at $\omega_c \tau \gg 1$ the dc current flows mainly in the strip of the width equal to the cyclotron diameter $l_c = 2l_0/(\omega_c \tau)$, which at these conditions is much less than the mean free path $l_0$. By contrast, at zero magnetic field the current is distributed over a much larger length $\sim l_0$.

\section{Summary}
We reported on the observation and study of cyclotron resonance induced edge photocurrents in high mobility hBN encapsulated bilayer graphene. 
%
%Unlike the case of zero magnetic field, the direction of edge photocurrents is only determined by the sign of the magnetic field and not by the radiation polarization. 
%
The cyclotron masses, momentum relaxation time, and carrier densities determined from the position and shape of CR together with the analysis of $1/B$-periodic magnetooscillations demonstrate that in studied samples and for the investigated carrier density range these parameters do not distinguish from those of the material bulk. 
Experimental findings for high carrier densities corresponding to the semiclassical regime are explained within the microscopic model based on the Boltzmann kinetic equation. Theoretical model describing the edge photocurrent at low carrier density, when the cyclotron resonance emerges due to inter-band or intra-band transitions between the lowest Landau levels, is a future task.

\section{Appendix 1}
\begin{figure}[h]
	\centering
	\includegraphics[width=\linewidth]{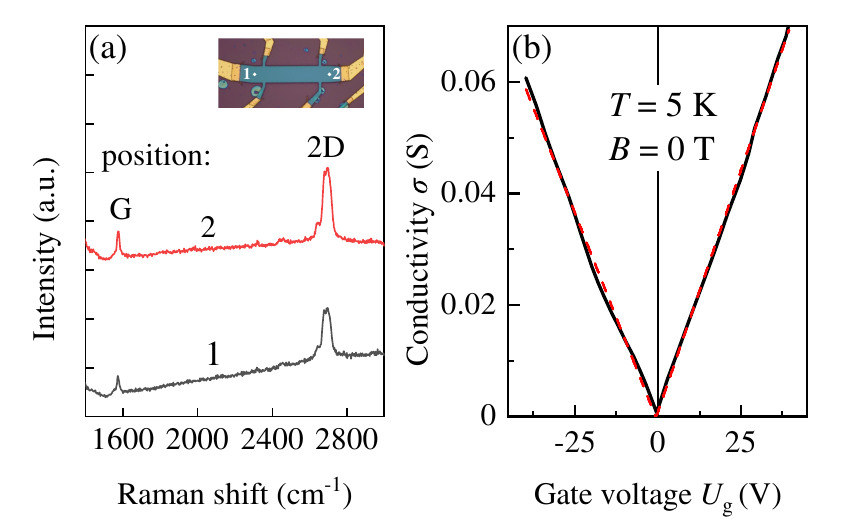}
	\caption{(a) Raman spectra of the final device. The measurements were carried out using a laser of $\lambda=532$~nm wavelength with the grating of 1200 gr/mm and the acquisition time of 2 min. The inset shows the sample micrograph, where the measured positions are marked with dots numbered 1 and 2. Panel (b) shows the dependence of the longitudinal conductivity on the gate voltage, measured at $T = 5$ K and $B = 0$ T. From the  slopes of the linear fits the mobilities were extracted:  $\mu_h\approx1.3\times 10^5 \text{cm}^2/\text{Vs}$ for the hole side and $\mu_e\approx1.4\times 10^5\text{cm}^2/\text{Vs}$ for the electron side. }
	\label{raman}
\end{figure}

\begin{figure}[h]
	\centering
	\includegraphics[width=\linewidth]{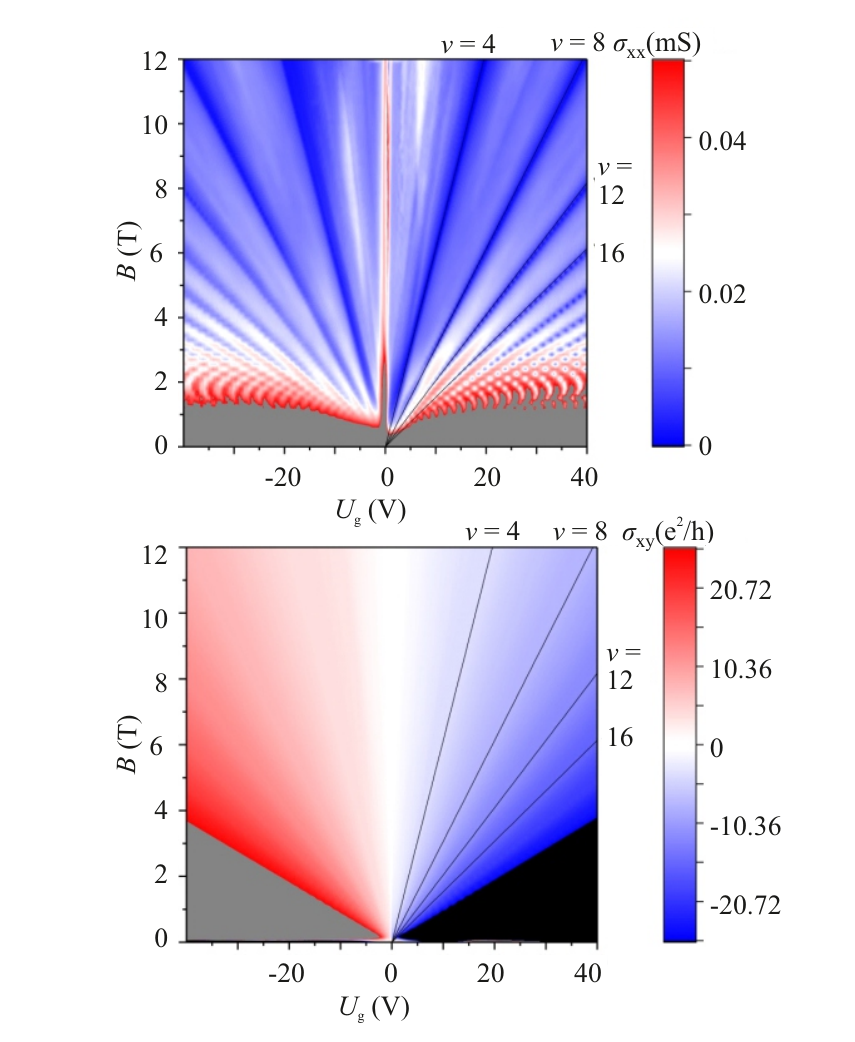}
	\caption{Magnetotransport data demonstrating the quantum Hall effect in our sample. Longitudinal (top panel) and transverse (bottom panel) conductivities are plotted as function of gate voltage and magnetic field. $T = 5$ K. Note that the sequence of filling factors ($\nu = \pm 4, \,\pm 8, \,\pm 12, \ldots$) follows that of the bilayer graphene spectrum, thus providing independent confirmation of the bilayer nature of our graphene sample.} 
	\label{transport_supplm}
\end{figure}

Figures~\ref{raman}~(a),(b), and \ref{transport_supplm} show the Raman spectra, dependence of the conductivity on the gate voltage, and magnetotransport results, respectively.

The micro-Raman scattering data obtained for two laser spot positions, see Fig.~\ref{raman}~(a) ensure that the  fabricated sample is bilayer graphene. The transport results are used to determine the mobility which is on the hole side $\mu \approx 1.3\times 10^5$~cm$^2$/Vs and on the electron side $\mu \approx 1.5\times 10^5$~cm$^2$/Vs. Note that the Drude charge carrier mobility approach cannot be used for the region close to CNP. From low temperature magnetotransport we can unambiguously confirm that our sample is the bilayer graphene. Namely, the longitudinal conductivity map shows strong cyclotron gaps (seen as dark blue lines in the top panel of Fig. \ref{transport_supplm}) at filling factors $\nu = n h / e B = \pm 4, \,\,\,\, \pm 8, \,\,\,\, \pm 12, \ldots$, consistent with the Landau level spectrum of bilayer graphene \cite{Novoselov2006, McCannFalko06}.

\section{Acknowledgments}
\label{acknow}

We thank S.A. Tarasenko,  I.A. Dmitriev, and H.~Plank for helpful discussions. The support from the Deutsche Forschungsgemeinschaft (DFG, German Research Foundation) – Project SPP 2244 (GA501/17-1), the Elite Network of Bavaria (K-NW-2013-247), and the Volkswagen Stiftung Program (97738) is gratefully acknowledged. S.D.G. acknowledges support of the Foundation for Polish Science (IRA Program, grant MAB/2018/9, CENTERA).  M.\,V.\,D. acknowledges financial support from the Russian Science Foundation (Project No. 19-72-00029).
VF and SS acknowledges support from
European Graphene Flagship Project (Core3), ERC Synergy
Grant Hetero2D, EPSRC grant EP/V007033/1.
AM acknowledges support of the EPSRC Early Career Fellowship (EP/N007131/1)and the European Research Council (ERC) under the European Union’s Horizon 2020 research and innovation programme (grant agreement No. 865590).

\bibliography{bilayer_bib}

\end{document}